# Aluminum lactate role in improving hydration and drying behavior of MgO-bonded refractory castables

D. S. Fini[(1)], V. C. Miguel[(1)], V. S. Pinto[(1)], V. C. Pandolfelli[(1)], M. H. Moreira[(2)], A. P. Luz[(1),*]

[(1)] Federal University of São Carlos, Materials Engineering Department (DEMa),
[(2)] Federal University of Sao Carlos, Graduate Program in Materials Science and Engineering (PPGCEM),
Rod. Washington Luiz, km 235, São Carlos, SP, 13565-905, Brazil.

*Corresponding author at: t*el.:* +55-16-3351-8601
E-mail: anapaula.light@gmail.com or analuz@ufscar.br

**Abstract**

Developing MgO-bonded castables is still an important subject for refractory producers and end-users based on the expansive character of the *in-situ* $Mg(OH)_2$ formation. Considering that magnesia undergoes hydration when exposed to water and the generated hydrated phase needs to be properly accommodated in the resulting microstructure to inhibit the generation of cracks, it is very important to find out alternatives to control/change the MgO hydration reaction rate, which may help to optimize the permeability and green mechanical strength of the castables. Therefore, fast and safer drying of such refractories can be carried out when adjusting $Mg(OH)_2$ generation with proper additives. This research investigated the use of aluminum lactate (AL) as a likely additive to change the hydration and drying behavior of vibratable castables bonded with different MgO sources (dead burnt, caustic or fumed one). Firstly, XRD, TG and DSC measurements of magnesia-based aqueous suspensions were evaluated to identify the AL effect on changing the hydration reaction products during the curing and drying steps. After that, $Al_2O_3$-MgO refractories were prepared and their flowability, curing behavior, cold flexural strength, apparent porosity, permeability and explosion resistance were evaluated. The results indicated that, instead of $Mg(OH)_2$, $Mg_6Al_2(OH)_{16}(OH)_2.4.5H_2O$/ $Mg_6Al_2(OH)_{16}(CO_3).4H_2O$ was the main hydrated phase identified in the AL-containing compositions. Due to this change in the hydration behavior of the refractories, the mixtures prepared with dead-burnt or magnesia fumes plus organic salt presented a longer setting time. Besides that, crack-free samples with improved permeability and green mechanical strength could be obtained when adding 0.5 wt.% or 1.0 wt.% of aluminum lactate to the tested castable compositions. Consequently, 1.0 wt.% of the selected additive favored the design of refractories with enhanced properties and greater spalling resistance, as no explosion could be observed even when subjecting the prepared samples to severe heating conditions (20°C/min).

1. Introduction

The development of MgO-bonded castables has been the subject of many investigations over the years [1–8]. Significant interest in this topic is due to challenges associated with the marked hydration likelihood of this oxide, leading to *in situ* $Mg(OH)_2$ formation, which is still one of the main obstacles to a more prominent use of magnesia in monolithic refractories. Magnesium hydroxide or the brucite phase is usually generated after a reaction of MgO with water in its liquid state or as vapor, during the castables' mixing/curing or drying steps, respectively [3]. When magnesia hydration takes place in the castable structure, the generated hydroxide crystals fill in the available pores and may act as a binder agent, leading to the hardening of the composition. However, this phase transformation must be followed carefully as it results in a 2.5-fold expansion due to the density mismatch between magnesia ($\rho = 3.5$ g/cm$^3$) and brucite ($\rho = 2.4$ g/cm$^3$), which usually gives rise to cracks in the consolidated refractory pieces with the growth of the formed $Mg(OH)_2$ crystals [1,9].

Among the various parameters that can affect MgO hydration during the castables' processing steps, the magnesia source and its interaction with other compounds are pointed out as the two most important aspects. Variations of this oxide can be obtained when thermally treating magnesite ($MgCO_3$) at different temperature ranges: caustic or calcined (CM), dead-burnt (DBM) and electrofused ones are produced at ~900-1300°C, ~1500-2000°C and above 2800°C, respectively. Depending on the production route, larger magnesia crystals (with a lower specific surface area and reduced hydration susceptibility [9–11], as indicated in Fig. 1) can be obtained. Nevertheless, if MgO hydration is considered as a binding alternative for castable compositions, fused magnesia is not a suitable material because it will not induce an effective brucite generation. On the other hand, the high reactivity of caustic MgO may generate an excessive amount of $Mg(OH)_2$ during water mixing and curing processes. Moreover, a more reactive magnesia may also induce faster *in situ* spinel ($MgAl_2O_4$) formation in alumina-based formulations at high temperatures [9,12]. Consequently, it is of utmost importance to find solutions to adjust MgO hydration features so that different sources of this oxide can be used as alternative binder additives for monolithic refractories.

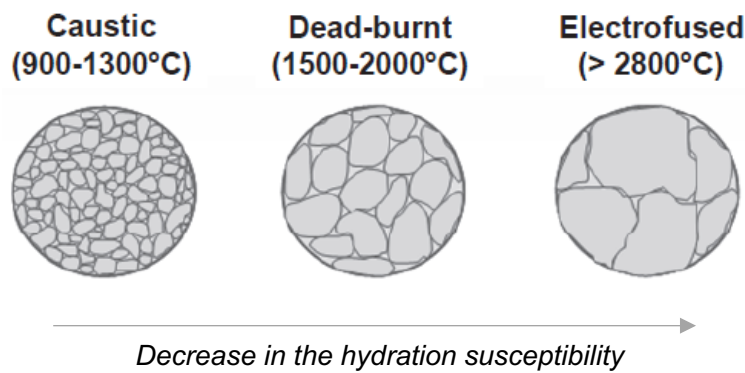

Fig. 1 – Effect of the primary crystal size on the hydration susceptibility of different magnesia sources (adapted from [9–11]).

Different strategies can be considered for having a better control of $Mg(OH)_2$ formation during the refractory processing steps, such as: *(i)* adding organic compounds (acetate, acids, etc.) to the prepared mixtures in order to change the morphology of the formed brucite crystals to better fit their growth in the resulting microstructure [13,14], and *(ii)* using hydrating agents (i.e., carboxylic acids) to favor a faster $Mg(OH)_2$ nucleation rate on a magnesia surface, which can limit its crystal growth [8,15]. Both alternatives focus on inducing brucite formation in the developed formulations, instead of inhibiting the hydration reaction [16,17], because the fresh molded MgO-bonded castables could still present enough room to accommodate stresses before their setting time. Thus, in such cases, it is desired to generate the hydroxide phase in the refractory microstructure in order to take advantage of its binding action.

Among the various organic additives suggested to improve the properties of cured magnesia containing compositions, aluminum lactate (aluminum salt of 2-hydroxypropanoic acid) is an interesting option as it can act as a chelating agent of $Mg^{2+}$ ions and induce the generation of hydrotalcite-like compounds $[Mg_xAl_y(OH)_{2x+2y}](CO_3)_{y/2}.nH_2O$ in such systems, which might lead to crack-free refractory pieces [15,18–22]. Previous works [23,24] also identified an additional benefit of incorporating this organic compound into magnesia-containing monolithic products. According to these studies, using a small amount of aluminum lactate in the designed compositions could favor the development of refractories with enhanced green mechanical strength, higher permeability and greater explosion resistance during their first heating treatment [23,24]. However, the setting time of the tested spinel-forming formulations was also reduced, which was pointed out as a negative effect of this salt addition [24]. Considering that the action mechanism of this additive was not clearly presented and discussed in the literature, further investigations are still required to understand the feasibility of incorporating this organic compound into formulations containing distinct MgO sources.

In general, it is desired that water withdrawal must take place as fast as possible in magnesia-containing compositions during their drying step. This is because hydrated phases are usually decomposed above 100°C and, due to the low porosity and permeability of the designed microstructures, the steam release is hindered. Consequently, vapor pressure builds up with the temperature and speeds up the magnesia hydration, causing damage to the refractories due to the destructive expansion mechanism [25]. Additionally, if MgO is exposed to steam for long periods of time, it will be more prone to rehydration. Considering that the magnesia particles derived from brucite decomposition (~380-420°C [15,26]) should have contact with water vapor during heating, this processing step is very critical due to the higher reactivity of this oxide.

Based on the presented aspects, this work addresses the investigation of $Al_2O_3$-MgO castable compositions containing distinct magnesia sources (dead-burnt, caustic or fume one, the latter is derived from the production of electrofused MgO) as binder components. Refractory formulations containing 0 - 1.0 wt.% of aluminum lactate were tested and their properties were analyzed throughout the curing and drying processing steps (30 up to 400°C), in order to identify the role of this organic salt in improving the castables' performance and adjusting the permeability level of the designed formulations.

**2. Materials and Methods**

*2.1 – Evaluation of aluminum lactate interaction with different MgO sources*

The following raw materials were evaluated in this study:

- *Aluminum lactate (AL)*: aluminum salt of 2-hydroxypropanoic acid (PA, 294.18 g/mol, Quimibras Indústrias Químicas S.A., Brazil);
- *Dead burnt magnesia (DBM)*: M-30B, MgO = 98.12 wt.%, CaO = 0.80 wt.%, $SiO_2$ = 0.34 wt.% and others = 0.74 wt.%, $d_{50}$ = 15.22 μm, RHI-Magnesita, Contagem, Brazil.
- *Caustic magnesia (CM)*: QMag 200AR, MgO = 98.27 wt.%, CaO = 0.82 wt.%, $SiO_2$ = 0.19 wt.% and others = 0.72 wt.%, $d_{50}$ = 12.77 μm, RHI-Magnesita, Contagem, Brazil.
- *Magnesia fumes (MF)*: MgO = 97.34 wt.%, CaO = 0.61 wt.%, $SiO_2$ = 0.64 wt.% and others = 1.41 wt.%, $d_{50}$ = 0.80 μm, RHI-Magnesita, Contagem, Brazil. This material is a very fine and reactive powder obtained during the production process of electrofused MgO.

The amount of the drying agent (AL) was defined based on preliminary tests carried out with magnesia-containing castable compositions, as the samples' cracking (due to brucite formation) could be inhibited when incorporating 6 wt.% of MgO and 1 wt.% of LA into the mixtures. For this reason, the same MgO:AL mass ratio (6:1) was kept when analyzing the simplified ceramic suspensions. Reference compositions (MgO + distilled water) were analyzed and compared to the ones containing 14.28 wt.% of AL (Table 1).

Table 1 – Magnesia-based suspensions prepared with and without AL for the evaluation of the likely chemical reactions that could take place between these materials.

| Mixtures | Prepared compositions (wt.%) | | Distilled water (wt.%) | pH |
|---|---|---|---|---|
| | MgO | Aluminum lactate (AL) | | |
| DBM | 100.00 | - | 30 | 10.97 |
| DBM-AL | 85.71 | 14.28 | 30 | 8.09 |
| CM | 100.00 | - | 40 | 10.04 |
| CM-AL | 85.71 | 14.28 | 40 | 9.16 |
| MF | 100.00 | - | 30 | 10.41 |
| MF-AL | 85.71 | 14.28 | 30 | 8.40 |

The dry powders (plain MgO or MgO + additive) were mixed with distilled water in a laboratory stirrer for approximately 5 minutes at room temperature. After that, pH measurements were carried out using a FiveEasy plus pH meter (Mettler Toledo, Brazil).

The prepared suspensions were cast as cylindrical samples (35 mm x 35 mm) and maintained at 30°C or 110°C for 24 hours. The obtained solid specimens were demolded, ground and evaluated via thermogravimetric (TG) and differential scanning calorimetry (DSC) measurements in an STA 449 F3 Jupiter device (Netzsch, Germany) in the 30-600°C temperature range, using 5°C/min as heating rate, synthetic air (80% $N_2$-20% $O_2$) flow of 50 cm$^3$/min and α-$Al_2O_3$ as a correction standard. X ray diffraction tests were also carried out with a D8 Focus equipment (Bruker, Germany), using CuKα radiation [λ=1.5418 Å], nickel filter, 40 mA, 40 mV, 2θ = 4-80° and scanning step = 0.02.

*2.2 – Design and characterization of vibratable $Al_2O_3$-MgO refractory castables*

The following stage of this study consisted of evaluating the effects of adding aluminum lactate to non-cement containing (NCC) MgO-bonded refractory castables. Six vibratable formulations (Table 2) were designed according to the Andreasen's model [9] and considering a distribution modulus ($q$) equal to 0.26, aiming to obtain suitable particle packing and define the appropriate amount of each selected raw material.

As highlighted in Table 2, in addition to the three magnesia sources and AL mentioned in Section 2.1, tabular alumina coarse aggregates (d < 6 mm, Almatis, Germany), as well as reactive and calcined aluminas (CT3000SG and CL370, Almatis, Germany) were the main materials added to the refractory compositions.

Table 2 – Vibratable castable compositions based on Andreasen's packing model (q = 0.26).

| Raw materials (wt.%) | Castable compositions | | | | | |
|---|---|---|---|---|---|---|
| | DBM-0AL | DBM-0.5AL | CM-0AL | CM-1AL | MF-0AL | MF-1AL |
| Tabular alumina (d < 6 mm) | 83 | 83 | 83 | 83 | 83 | 83 |
| Reactive alumina (CT3000SG) | 7 | 7 | 7 | 7 | 7 | 7 |
| Calcined alumina (CL370) | 4 | 4 | 4 | 4 | 4 | 4 |
| Dead-burnt MgO (DBM) | 6 | 6 | - | - | - | - |
| Caustic MgO (CM) | - | - | 6 | 6 | - | - |
| MgO fumes (MF) | - | - | - | - | 6 | 6 |
| Aluminum lactate (AL) | - | 0.5 | - | 1 | - | 1 |
| Dispersant (Castament® FS 60) | 0.2 | 0.2 | 0.2 | 0.2 | 0.2 | 0.2 |
| Distilled water | 4.2 | 4.2 | 5.2 | 5.2 | 4.2 | 4.2 |

The amount of AL used for each magnesia source was defined based on preliminary tests, as well as previous works presented in the literature [8,27]. During the mixing step, a total of 0.2 wt.% of a polyethylene glycol-based dispersant (Castament® FS 60, Basf, Germany) was incorporated into the dry-powders of the formulations and their homogenization was carried out in a rheometer device [28]. The water demand for each tested composition was adjusted in order to obtain vibratable flow values of approximately 150% (ASTM C 1445).

In a further step, the prepared mixtures were cast, cured at 30°C for 24 hours and dried at 110°C for more 24 hours. The curing behavior and setting time of the fresh mixtures were followed via ultrasonic measurements (UltraTest device, IP-8 measuring system, Germany) at room temperature (22°C) for 24 hours, in order to evaluate the propagation velocity of ultrasonic waves in the prepared materials as a function of time.

Thermogravimetric tests were conducted in an electric furnace controlled by a proportional-integral-derivative (PID) system up to a maximum temperature of 600°C and using heating rates of 5 or 20°C/min. Cylindrical samples (50 mm x 50 mm), previously cured at 30°C for 24 hours or cured at 30°C and dried at 110°C for more 24 hours had their mass loss and inner temperature monitored during each heating schedule. Further details of the device used during these tests can be found elsewhere [29]. The highest heating rate (20°C/min) was chosen to analyze the castables' explosion likelihood as this is a very severe condition and most industrial drying schedules are only based on very conservative rates (i.e., 100°C/h).

Moreover, mass loss during drying was measured based on the cumulative water content expelled during heating per total amount of water initially contained in the humid body [29,30]. The derivative of the mass loss profiles as a function of time was also calculated using the Origin software (version 9, OriginLab, USA).

Cold mechanical strength was analyzed via 3-point bending tests (ASTM C133-97) on cast prismatic samples (150 mm x 25 mm x 25 mm) cured at 30°C for 24 hours and dried at 110°C for 24 hours using MTS-810 equipment (Material Test System, USA). Additionally, the apparent porosity of such materials was calculated according to ASTM C380-00 and using kerosene as immersion liquid. A set of five samples was evaluated for each testing condition and the presented values are the average result with their respective standard deviation.

Besides that, air permeability measurements were carried out at room temperature in a permeameter device developed for this purpose (further details of the used equipment can be found elsewhere [30]). The evaluated castable samples consisted of cylinders with 70 mm of diameter and 25 mm of height. Such materials were cured at 30°C for 24 hours, dried at 110°C for 24h and/or calcined at 250 or 450°C for 5 hours. The samples were held between two chambers, leaving a circular passing area of 38.2 mm$^2$. The inlet air pressure ($P_i$) and volumetric air-flow rates (Q) were monitored during the measurements performed under steady-state conditions. Additionally, the inlet and outlet ($P_0$) pressures were collected using an electronic transducer and a U-type water manometer.

To calculate the Darcian ($k_1$) and non-Darcian ($k_2$) permeability constants of the designed formulations containing aluminum lactate, the Forchheimer's equation (Eq. 1) was considered [31]:

$$\frac{P_i^2 - P_0^2}{2P_0 L} = \frac{\mu}{k_1} V_s + \frac{\rho}{k_2} V_s^2 \tag{1}$$

where $L$ (m) is the sample's thickness, $V_s$ (m/s) is the air velocity, $\mu$ is the air viscosity (1.8 x 10$^{-5}$ Pa.s) and $\rho$ is the air density (1.08 kg/m$^3$) at room temperature. Four samples of each selected composition and temperature (110, 250 or 450°C) were analyzed in these tests.

## 3. Results and Discussion

*3.1 – Aluminum lactate effects on MgO hydration behavior*

As expected, periclase and brucite [Mg(OH)$_2$] were the only crystalline phases detected in the additive-free compositions (Fig. 2a). Moreover, the peaks' intensity of the hydroxide compound increased as a function of the reactivity of the tested MgO (CM > MF > DBM), as they presented distinct specific surface area: CM = 21.31 m$^2$/g, MF = 5.6 m$^2$/g and DBM = 0.56 m$^2$/g. Thus, the higher hydration susceptibility of caustic magnesia when in contact with liquid water at 30°C was confirmed, which could result in an enhanced binding action of this oxide when added to refractory compositions.

Nevertheless, the higher reactivity of CM can also lead to the development of cracked samples, due to the non-accommodation of the greater amount of brucite crystals in the resulting microstructure. Thus, it was important to analyze the addition of aluminum lactate to the MgO-suspensions to identify how this compound could affect the likely phase transformations in conditions similar to those used during the processing of refractory compositions. Fig. 2b presents the diffractograms collected for the aluminum lactate (AL)-containing suspensions.

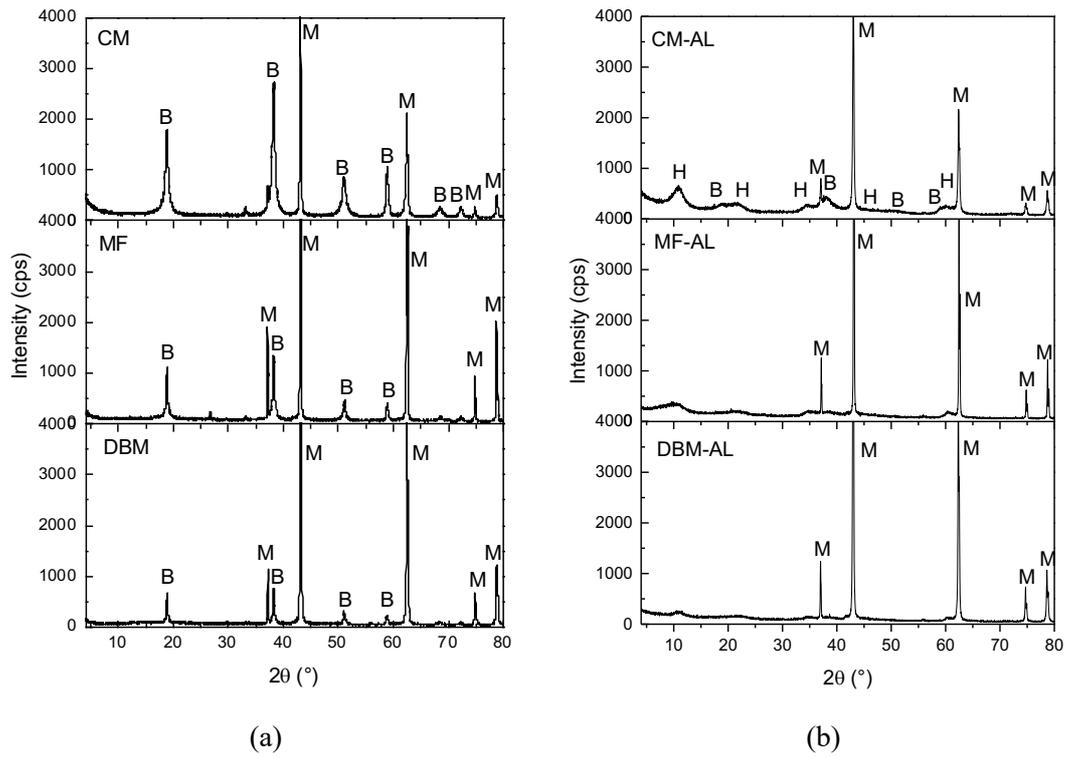

Fig. 2 – XRD profiles of the prepared MgO-based suspensions after curing the samples at 30°C for 24 hours. (a) Plain magnesia (DBM = dead burnt, FM = fumes, CM = caustic) and (b) magnesia plus aluminum lactate (AL). Identified phases: M = periclase (MgO); B = brucite [$Mg(OH)_2$]; H = hydrotalcite-like phase [$Mg_6Al_2(OH)_{16}(OH)_2 \cdot 4.5H_2O$/ $Mg_6Al_2(OH)_{16}(CO_3) \cdot 4H_2O$].

Periclase was the main crystalline phase identified in Fig. 2b and only the CM-AL sample presented additional broad peaks that could be associated with hydrotalcite-like phases [$Mg_6Al_2(OH)_{16}(OH)_2 \cdot 4.5H_2O$ / $Mg_6Al_2(OH)_{16}(CO_3) \cdot 4H_2O$], resulting from the interaction of the aluminum salt with this more reactive magnesia and water. Both suggested phases consist of layered double hydroxides containing metal cations ($M^{2+}$ and $M^{3+}$) of similar radii, which are randomly distributed in the octahedral positions, resulting in brucite-like structures [32]. Usually, the brucite-type layers are stacked on top of each other and are held together by weak hydrogen bonds (Fig. 3). The ratio of $M^{2+}$ and $M^{3+}$ cations determines the mineral structure and, to maintain electroneutrality, the interlamellar domain must be occupied by a suitable number of anions (generally hydrated ones such as hydroxides, anionic complexes, organic anions, etc.) [18,32].

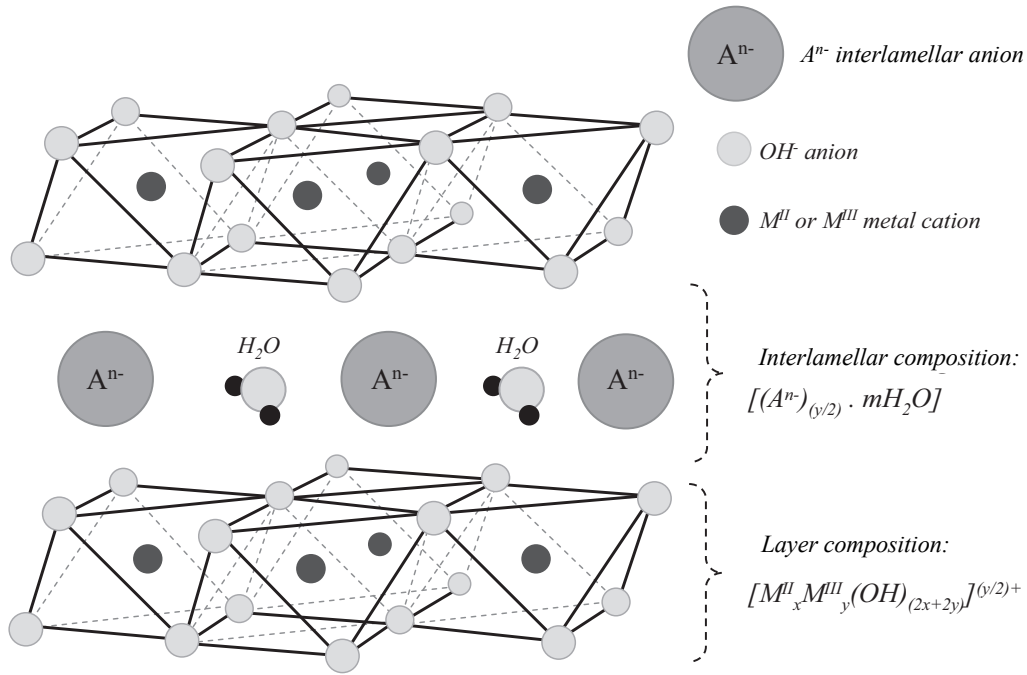

Fig. 3 – Schematic representation of hydrotalcite structure [18].

Many types of hydrotalcite can be formed from different combinations of divalent and trivalent cations and distinct interlamellar anions. The X ray diffraction peaks identified in Fig. 2b can be related to two hydrotalcite-like phases (hydrotalcite and meixnerite, Table 3), as they present similar diffraction patterns, crystallographic parameters and symmetry, which makes it difficult to detect the contribution of each phase.

Table 3 – Compositions, crystallographic parameters and symmetries for some natural layered double hydroxide phases [32].

| Name | Chemical composition | Unit cell parameters | | Symmetry |
|---|---|---|---|---|
| | | a (nm) | c (nm) | |
| Hydrotalcite | $Mg_6Al_2(OH)_{16}CO_3 \cdot 4H_2O$ | 0.3054 | 2.281 | 3R |
| Meixnerite | $Mg_6Al_2(OH)_{16}(OH)_2 \cdot 4H_2O$ | 0.3046 | 2.292 | 3R |

According to the XRD results (Fig. 2), AL addition to the MgO-suspensions inhibited brucite formation. Hence, the main benefit of the tested organic salt was the generation of the layered double hydroxides on the magnesia particles' surface, which might have prevented water from diffusing due to the low solubility of hydrotalcite-like compounds in an alkaline environment and, consequently, limited MgO hydration [33].

When analyzing the MgO-based samples maintained at 110°C for 24 hours (results not shown here), the same phases contained in the cured materials were identified and a small increase of their peaks' intensity could be observed. The partial crystallization of $Mg_6Al_2(OH)_{16}(OH)_2 \cdot 4.5H_2O$/$Mg_6Al_2(OH)_{16}(CO_3) \cdot 4H_2O$ and brucite have been favored in such condition in all AL-containing compositions, as the shape of their respective diffraction peaks were better defined after drying the samples at 110°C.

According to Fig. 4a and 4b, the mass loss and DTG profiles of the plain MgO suspensions kept at 30°C for 24 hours pointed out that two main transformations took place when heating the samples up to 600°C. Firstly, mass loss due to free-water release was identified close to 80-100°C and, later brucite decomposition could also be observed in the range of 350-400°C. Greater mass loss and more significant DTG peaks were obtained for CM suspension (Fig. 4a and 4b), which can be attributed to the higher amount of water required during its preparation (Table 1).

In the case of the AL-containing compositions, the same transformations described above could still be detected, but some changes in the mass loss profiles might be associated with the hydrotalcite-like phase presence. For instance, some authors reported that the decomposition of the Mg,Al hydrotalcite structure should occur in three steps [15,32]: *(i)* removal of free or adsorbed water (< 100°C), *(ii)* elimination of the interlamellar structural water (100-200°C), and *(iii)* the simultaneous dehydroxylation and decarbonation of the layered double hydroxide framework (300-400°C). Hence, the initial peak of the AL-containing samples shown in Fig. 4b represents both free-water and interlamellar structural water withdrawal, which took place in an extended temperature range (mainly 50-200°C). Moreover, an intense peak associated with hydrotalcite-like and brucite phases decomposition was identified around 300-400°C. Such results are in tune with the ones presented by Palmer [32] for the characterization of Mg-Al hydrotalcite.

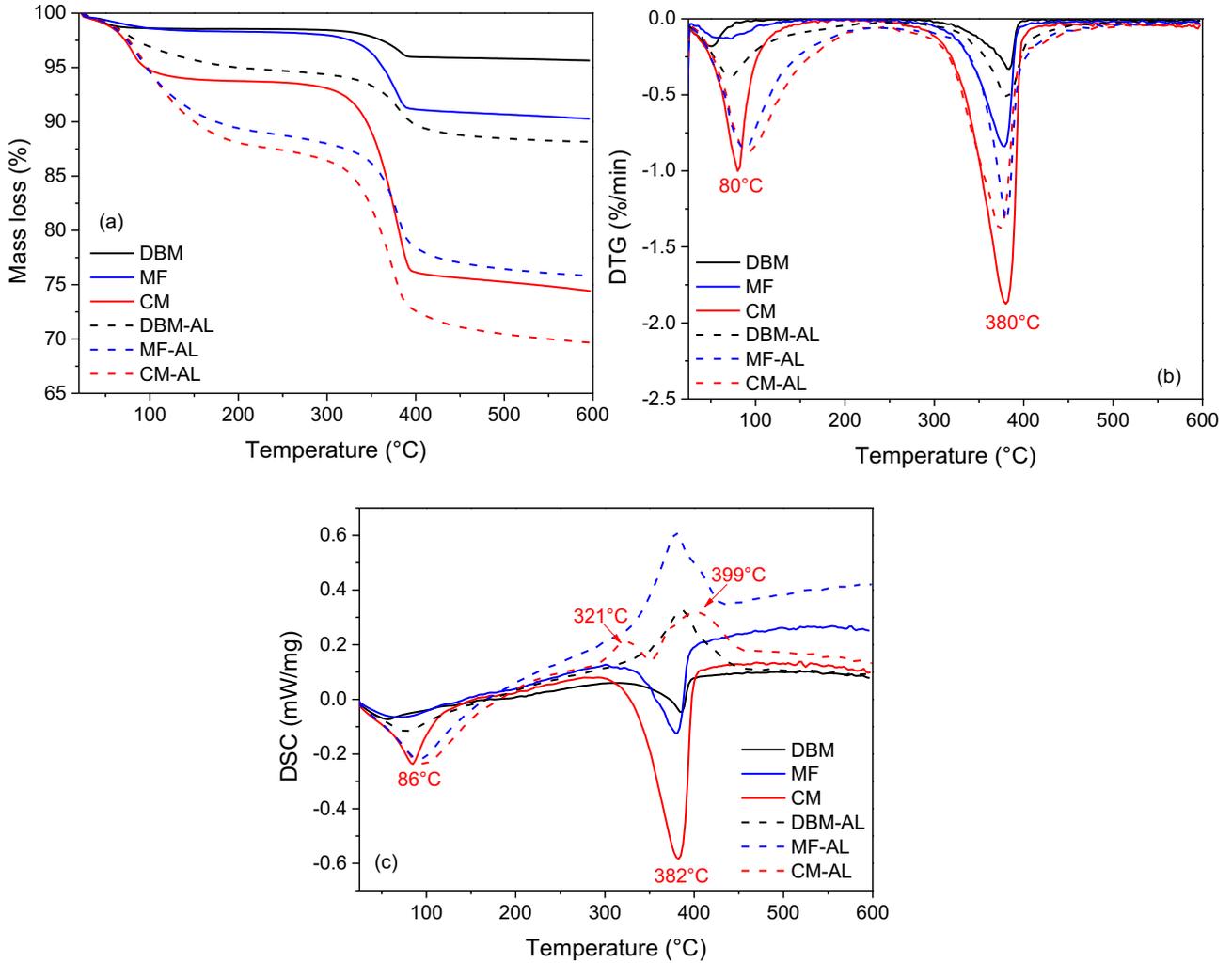

Fig. 4 – (a) Mass loss, (b) DTG and (c) DSC profiles of the MgO-based suspensions after curing step at 30°C for 24 hours.

Regarding the DSC profiles of the tested samples (Fig. 4c), some differences between plain MgO-suspensions and AL-containing ones could be observed mainly in the 300-400°C temperature range. The reference samples presented two endothermic peaks around 86°C and 382°C due to free-water release and brucite decomposition, respectively. On the other hand, instead of those two endothermic peaks, the incorporation of aluminum lactate into the mixtures resulted in samples showing an initial endothermic peak (related to the free/adsorbed and interlamellar water withdrawal) followed by one or two intense exothermic peaks between 300-450°C (Fig. 4c).

An additional test was carried out to analyze the thermal decomposition of plain aluminum lactate (as received). Fig. 5 indicates that the main mass loss of this additive took place around 375°C, which commonly induces the formation of amorphous $Al_2O_3$, $CO_{(g)}$ and $H_2O_{(g)}$ as main products (Eq. 2) [34,35]. Following this transformation, an exothermic reaction based on the interaction of steam with carbon monoxide may lead to the production of carbon dioxide and hydrogen gas, as shown in Eq. 3 [35]. Consequently, as stated by Sato et al. [34], the thermal decomposition of $Al[CH_3CH(OH)COO]_3$ is based on the decomposition of the organic skeleton and the combustion of the formed products (water-gas shift reaction, Eq. 3). As a result of the latter transformation, a major increase of the DSC signal was identified in the 300-400°C range for aluminum lactate (Fig. 5b), which explains why the evaluation of the AL-containing samples resulted in exothermic peaks at the same temperatures (Fig. 4b), instead of presenting the expected endothermic transformations associated with brucite and hydrotalcite decomposition.

$$Al[CH_3CH(OH)COO]_3 \rightarrow amorphous\ Al_2O_3 + CO_{(g)} + H_2O_{(g)} \qquad (2)$$

$$H_2O_{(g)} + CO_{(g)} \leftrightarrow CO_{2(g)} + H_{2(g)} \qquad \Delta H = -40.6\ kJ/mol \qquad (3)$$

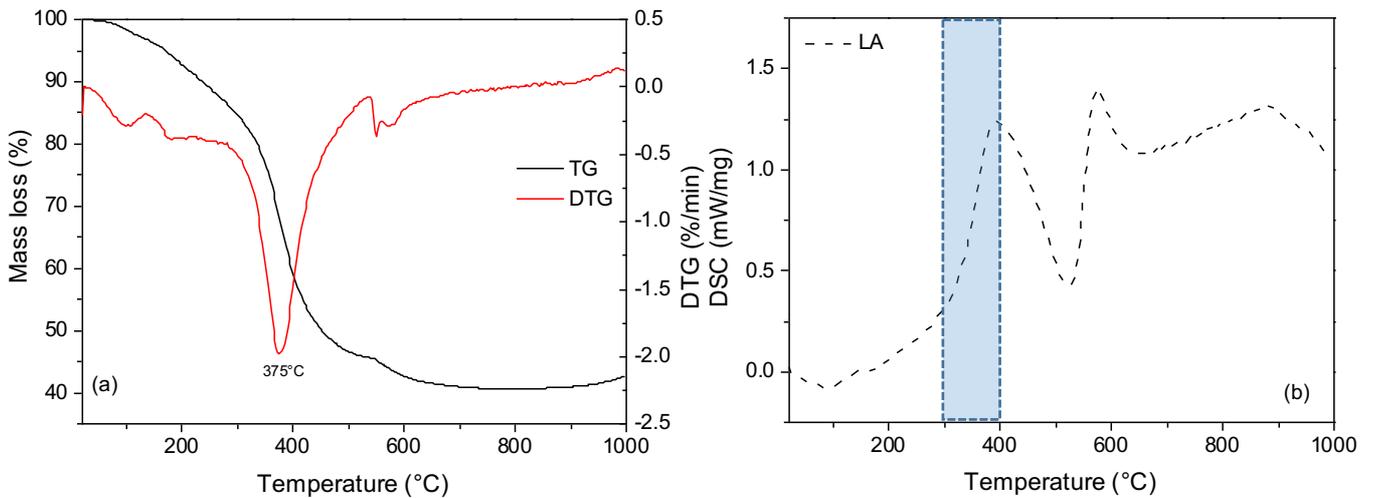

Fig. 5 – (a) Mass loss and DTG, and (b) DSC profiles of the aluminum lactate (AL) evaluated in this work. Blue rectangle highlights the temperature range where AL-based suspensions presented exothermic peaks (as shown in Fig. 5b).

When analyzing the dried samples (kept at 110°C for 24h) of the prepared suspensions via TG/DSC measurements, similar results to the ones presented in Fig. 4 were observed, except that the initial mass loss event between 80-200°C, related to free/adsorbed and interlamellar water, was not detected.

Based on the XRD and TG/DSC data presented before, it can be expected that the incorporation of AL into castables might lead to microstructural changes and affect their properties due to the modification of the magnesia hydration reaction products. In order to investigate this aspect, the following step of this work consisted of designing vibratable refractory castables containing this additive and distinct MgO sources.

*3.2 – Characterization of MgO-bonded refractory castables with and without aluminum lactate (AL)*

Preliminary tests indicated that 0.5 wt.% of aluminum lactate was required to be added to DBM-bonded compositions in order to prevent their cracking during the samples drying at 110°C, whereas the more reactive magnesia (CM and MF) demanded 1 wt.% of this compound. In general, the presence of the organic salt led to some small changes in the refractory flowability when comparing them to the AL-free mixtures (reference materials), as shown in Fig. 6a.

However, aluminum lactate played an important role in the hardening behavior of the prepared mixtures, which could be inferred by carrying out ultrasonic measurements of the fresh castables as a function of time. Fig. 6b indicates that as a result of magnesia hydration and brucite crystals' precipitation, DBM-0AL, CM-0AL and MF-0AL compositions presented a fast increase in the ultrasonic wave propagation velocity at the initial moments of their curing step at ~22°C.

These results confirm that the setting and hardening of the samples without AL took place between 2-5 hours and the CM-0AL formulation presented the highest velocity after 24 h, which suggests that a greater amount of brucite might be properly accommodated in the pores and voids of the resulting microstructure, leading to a more effective binding effect. It can also be highlighted that no cracking of the samples was detected when keeping them at 22°C for 24 hours, as no propagation velocity decay was observed in the curves shown in Fig. 6b.

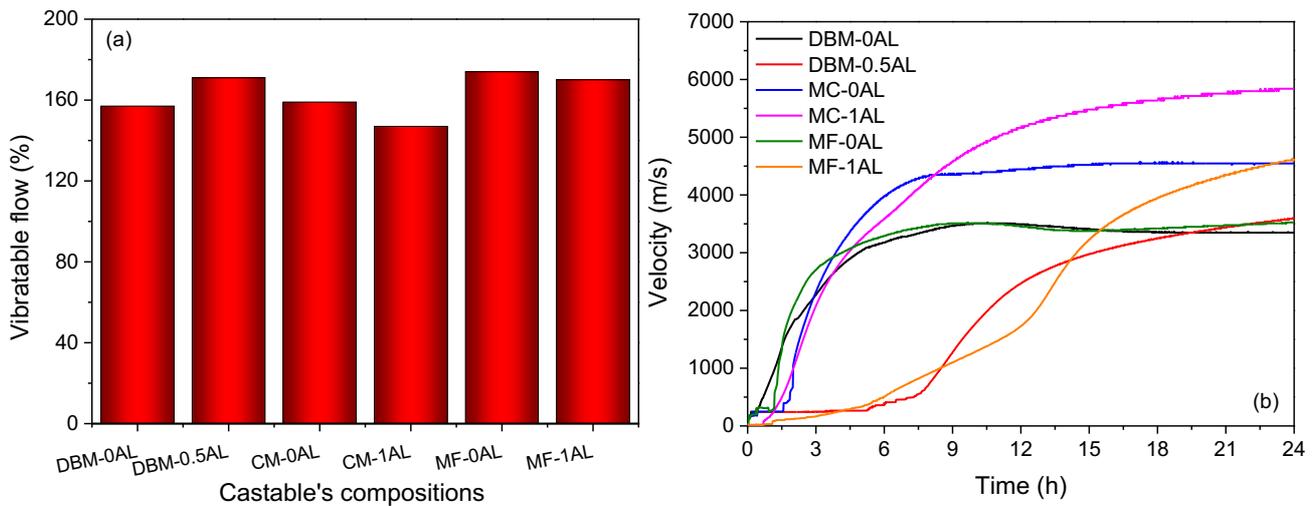

Fig. 6 – (a) Flowability and (b) ultrasonic propagation velocity as a function of time for the designed refractory castables.

Longer setting times were identified when testing DBM-1AL and MF-1AL (Fig. 6b, velocity increase took place only above 7 hours of curing) and such results are in agreement with the XRD measurements carried out for the MgO-based suspensions, as the precipitation of brucite crystals was inhibited when aluminum lactate was incorporated into the mix containing dead-burnt and magnesia fumes (Fig. 2b). The selected additive acted preventing MgO hydration and modifying the reactions sequence, resulting in the generation of hydrotalcite-like phases [$Mg_6Al_2(OH)_{16}(OH)_2.4.5H_2O$ / $Mg_6Al_2(OH)_{16}(CO_3).4H_2O$] and brucite only when exposing the DBM and MF-containing samples at higher temperatures. Consequently, these chemical changes were responsible for the distinct hardening behavior of the DMB-1AL and MF-1AL compositions. In the case of CM-1AL, most likely the higher reactivity of this magnesia source and greater water availability in the prepared mixture helped to maintain the setting time of this formulation similar to the one without AL (CM-0AL, Fig. 6b). Moreover, the precipitation of crystalline hydrates was identified even during the curing time (~30°C) in the CM-AL suspension (Fig. 2b).

Yeh et al. [24] reported that aluminum lactate added to a spinel-forming castable sped up its hardening time, which was highlighted as a drawback of this additive. However, such authors evaluated a calcium aluminate cement-bonded formulation also containing silica in addition to $Al_2O_3$ and MgO. Thus, it is very likely that the interaction of AL with other components (i.e., cement and/or silica) of the evaluated system influenced the obtained setting behavior.

Considering that the final ultrasonic wave velocity measured for CM-1AL sample was the highest one (Fig. 6b), it was expected that this castable should present an improved green mechanical strength when compared to the other compositions. Fig. 7a shows that this trend was confirmed in all tested conditions and, additionally, AL helped to inhibit the refractories' cracking due to the modification of the MgO hydration behavior, as discussed in Section 3.1.

The reference formulations (AL-free refractories) did not present any cracks on the samples' surface after keeping them at 30°C for 24h. However, a high number of cracks was observed in these materials after drying at 110°C for another 24h (see Fig. 7c). This performance indicates that brucite crystal precipitation and growth took place when exposing the prepared castables to an environment containing water vapor. The expansive feature of this phase transformation resulted in mechanical stresses in the formed microstructure and, as a consequence, cracks were generated and led to a major decrease of DMB-0AL, CM-0AL and MF-0AL flexural strength values (Fig. 7a). Due to the high number of cracks contained in the dried reference samples, their mechanical strength and apparent porosity after firing treatment at 250°C and 450°C, were not evaluated.

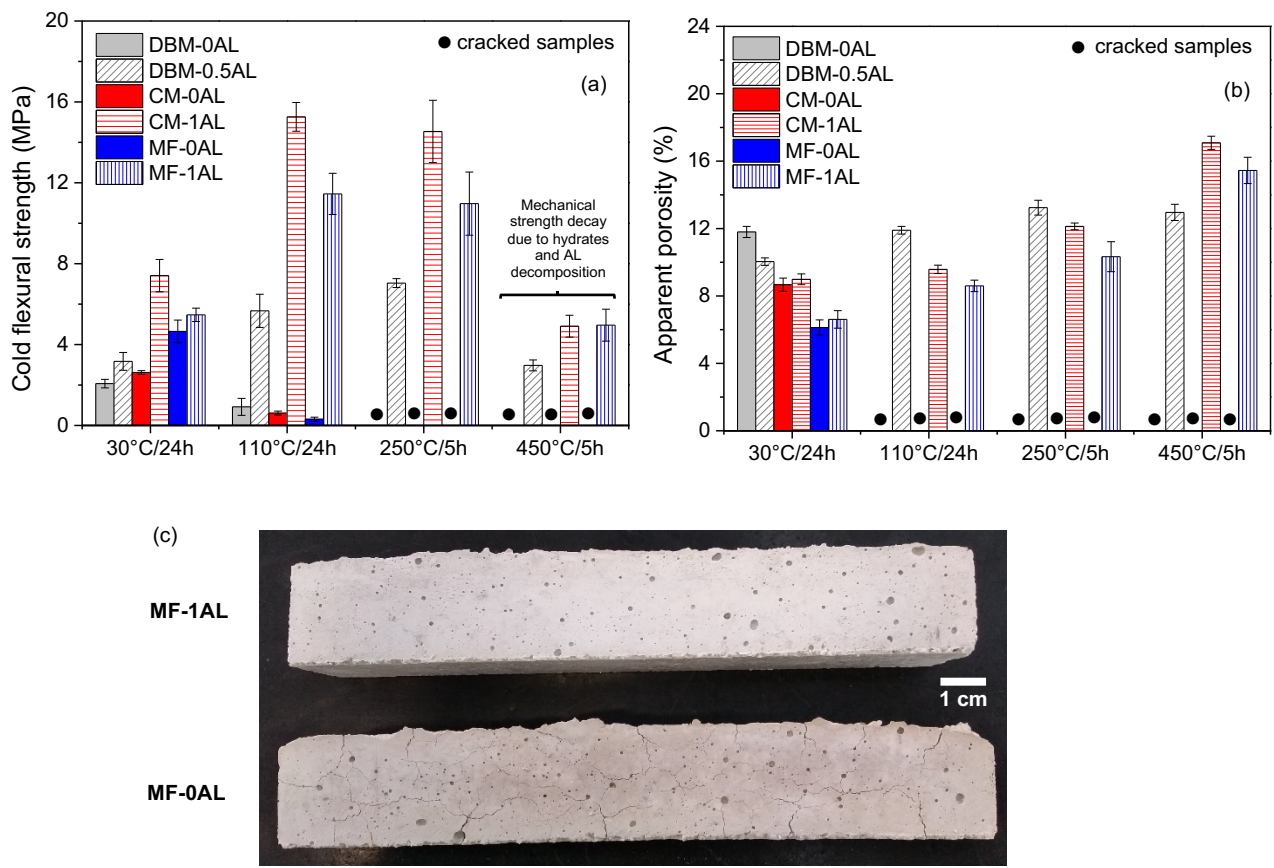

Fig. 7 – (a) Cold flexural strength and (b) apparent porosity of the evaluated castable compositions containing 0 - 1 wt.% of aluminum lactate (AL). (c) Images of the MF-bonded castable samples after drying at 110°C for 24h.

As observed in Fig. 7, two important changes in the AL-containing castables' properties were observed when adjusting the MgO hydration with the aid of the selected organic salt: *(i)* enhancement of the green mechanical strength and no crack formation due to the better accommodation of the hydrated phases in the microstructure, and *(ii)* continuous increase of the measured apparent porosity when increasing the firing temperature. Both aspects are important when thinking about designing refractories with optimized heat up explosion resistance, as the mechanical strength of the ceramic linings and their ability to release water by permeable paths will define the vapor pressure build up at a certain depth from the refractory surface [29,36]. Consequently, depending on the applied heating rate, the vapor formation rate can be very dangerous for the green castables, as it implies in pressure values that are high enough to lead to the material's explosion [9].

Due to the improved flexural strength of the CM-1AL castable in the 30-450°C range (Fig. 7a), it is expected that this refractory may withstand higher thermomechanical stresses when compared to other designed compositions. However, despite this positive aspect and the increased apparent porosity of the AL-containing formulations, the samples' permeability is another property that needs to be analyzed to infer the influence of aluminum lactate in changing the drying behavior of the MgO-bonded refractories.

The dry out process of castables bonded with hydraulic binders (i.e. MgO) usually involves three stages that take place at different temperatures (up to 550-600°C, depending on the refractories' permeability and applied heating rates): *(i)* evaporation, *(ii)* ebullition and *(iii)* hydrates decomposition. The relative amount of physical (free-water expelled during evaporation or ebullition) and chemically-bonded $H_2O$ depends on the total liquid and binder contents added during the castables mixing [1,9]. Ebullition is the most critical dewatering step and where most likely spalling takes place, as microstructures with reduced permeability will trap the formed vapor and result in pressure values that may be high enough to induce the ceramic explosion. As evaporation, ebullition and hydrates decomposition lead to changes in the refractory microstructure, it was important to identify the share of these changes in the permeability of the designed castables. Therefore, the AL-containing samples were dried at 110°C for 24h or pre-fired at 250°C and 450°C for 5h.

Fig. 8 shows the Darcian ($k_1$) and non-Darcian permeability ($k_2$) constants measured at room temperature after evaluating the prepared compositions. Considering the fact that the reference materials (AL-free formulations) cracked during drying at 110°C for 24h, only the castables bonded with different MgO sources and containing 0.5 or 1 wt.% of aluminum lactate were tested.

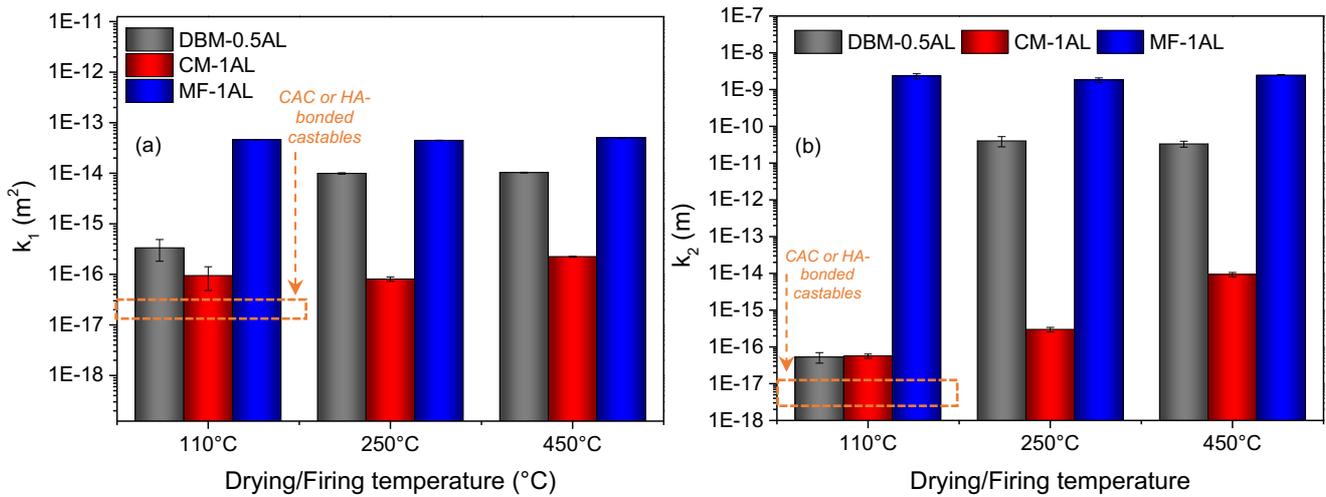

Fig. 8 – Permeability constants (a) $k_1$ and (b) $k_2$ measured for the dried and fired samples of the AL-containing castables.

In general, the CM-1AL castables presented the lowest $k_1$ (which is related to the energy losses due to viscous friction of the fluid with the refractory structure) and $k_2$ (denotes the share of inertia and turbulence on the pressure drop of the fluid) values in the analyzed conditions when compared to the other compositions containing DBM or MF as magnesia sources (Fig. 8). However, as highlighted in Fig. 8, the obtained permeability results were higher than the ones commonly observed for calcium aluminate cement (CAC) or hydratable alumina (HA)-bonded refractory systems [27,37]. Hence, AL addition to the designed compositions induced the generation of a greater number of permeable paths, which is a key aspect for safe and fast drying, as the castable's mechanical strength itself may not be enough to withstand the high thermomechanical stress.

MF-1AL was the only refractory that already presented high permeability values and minor changes in these results in the 110-450°C temperature range, whereas CM-1AL and DMB-1AL showed a continuous $k_1$ and $k_2$ increase with temperature (Fig. 8). The increase in the number of permeable paths of these refractories is most likely related to the release of adsorbed and interlamellar water (80-200°C) and brucite, hydrotalcite-like phase and aluminum lactate decomposition (300-420°C) during the samples' drying and firing steps.

Due to the critical damage that can occur during drying, the evaluation of this processing step is essential for MgO-bonded castables. The main technique applied to follow the brucite decomposition and the associated spalling effect is the thermogravimetric analysis (which can indicate the hydration degree) [9]. By measuring the percentage of mass loss (W, %) and by calculating the drying rate (DTG, %/min), it is possible to identify whether $Mg(OH)_2$ and $Mg_6Al_2(OH)_{16}(OH)_2.4.5H_2O$ / $Mg_6Al_2(OH)_{16}(CO_3).4H_2O$ were formed.

Fig. 9 shows the mass loss and DTG profiles obtained for cured (30°C for 24 h) and dried (110°C for 24h) samples. A large amount of liquid must be removed from the dense microstructure of the cured refractories during their first heating treatment, which makes this a critical condition resulting in samples that are more prone to explode. Fig. 9a pointed out that DBM-0AL and MF-0AL did not withstand the stresses derived from the pressure built up due to free-water release during the ebullition step (~ 148°C). CM-0AL, on the other hand, showed a major mass loss close to 132°C, indicating that the generated steam was successfully permeated from the inner region to the external surface of the samples without causing damage to this material. Consequently, the spalling of cured DBM-0AL and MF-0AL compositions is most likely related to the precipitation of the brucite phase during curing, giving rise to a microstructure with a reduced number of permeable paths for further gas withdrawal during drying. Besides that, additional $Mg(OH)_2$ precipitation during the samples' heating (due to the interaction of magnesia particles with water vapor) could take place and induce their cracking as pointed out in Fig. 7, which could also favor DBM-0AL and MF-0AL explosion.

As CM-0LA presented intermediate values of green mechanical strength and porosity (when comparing DBM-0LA and MF-0LA, Fig. 7), most likely its resulting microstructure presented a suitable number of permeable paths after curing. Additionally, the brucite phase formed during heating was properly accommodated in the samples, which did not lead to their cracking during the thermogravimetric measurements and made it possible for this castable to withstand the generated steam pressure.

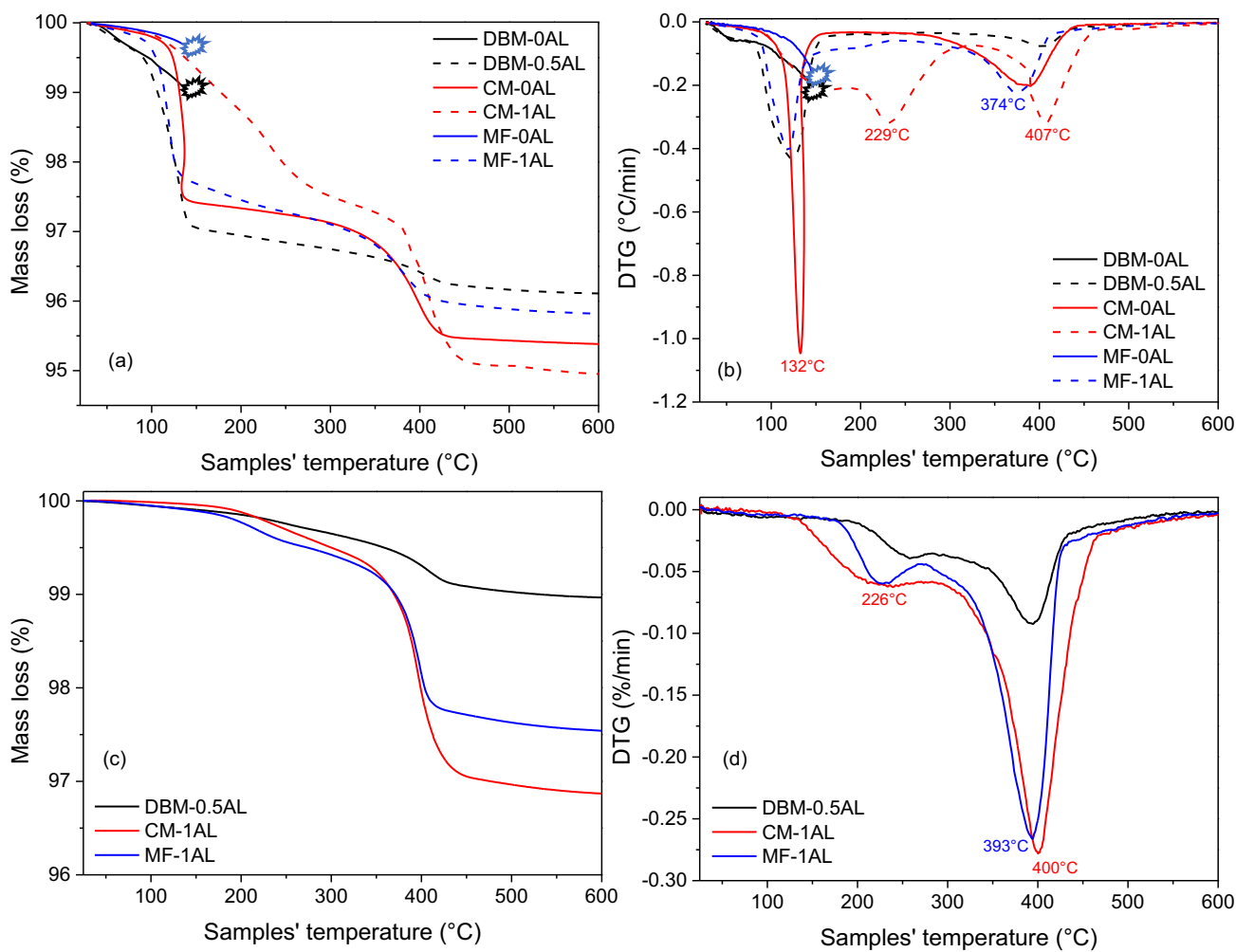

Fig. 9 – Mass loss and DTG profiles of the MgO-bonded castables after (a and b) curing 30°C for 24h and (c and d) drying at 110°C for 24h steps. All samples were subjected to a heating rate of 20°C/min.

Regarding the AL-containing castables, all cured samples did not explode during the heating step up to 600°C with a heating rate of 20°C/min (Fig. 9a and 9b). The main mass loss events for DBM-0.5AL and MF-1AL refractories took place around 120°C (free-water release) and 374-400°C (most likely AL decomposition, as a limited amount of hydrotalcite-like phases was formed in DBM-AL and MF-AL-based suspensions after 30°C for 24h, Fig. 2b). Conversely, $Mg(OH)_2$ and $Mg_6Al_2(OH)_{16}(OH)_2 \cdot 4.5H_2O$/$Mg_6Al_2(OH)_{16}(CO_3) \cdot 4H_2O$ was generated in CM-1AL during the curing step (Fig. 2b), resulting in a slower and continuous mass loss from 100°C to 300°C, as a consequence of the adsorbed and interlamellar water release (Fig. 9a). After that, the decomposition of these phases and AL is responsible for the further mass loss of the AL-containing castables in the range of 300-420°C.

When the prepared refractories were subjected to a drying step at 110°C for 24h before the TG measurements, it was observed that the initial mass loss around 80-100°C was suppressed in the obtained profiles (Fig. 9c). However, the other two events, peaks at 226°C and around 400°C (Fig 9d), were still detected due to the presence of the hydrotalcite-like phases, as well as brucite and unreacted aluminum lactate in such compositions. The reference refractories (AL-free compositions) did not have their spalling resistance evaluated after their drying step because the prepared samples presented a large number of cracks on their surface. This fact indicates that AL is an interesting additive, as it not only helps to inhibit MgO hydration, but also improves the explosion resistance of dense refractories containing different sources of this oxide.

## 4. Conclusions

The present work investigated the use of aluminum lactate (AL) as a likely additive to change the hydration and drying behavior of vibratable castables bonded with different MgO sources (dead burnt, caustic or fumed one).

According to the obtained results, the evaluation of magnesia-based suspensions attested that caustic magnesia was the most reactive MgO source tested in this study, which led to the formation of brucite and hydrotalcite-like phases [$Mg_6Al_2(OH)_{16}(OH)_2.4.5H_2O$/ $Mg_6Al_2(OH)_{16}(CO_3).4H_2O$] in the samples containing AL kept at a low temperature (30°C for 24h). The other tested magnesia (dead-burnt or fumed one) only showed the formation of these hydrates after the drying step at 110°C for 24h. The main benefit of aluminum lactate addition to the designed compositions was the generation of the layered double hydroxides on the magnesia particle's surface, which limited MgO hydration and brucite formation. As a consequence, MgO-bonded castables could be prepared without the development of cracks on their surface.

Besides that, AL affected the setting time (requiring longer time for the samples' hardening) of the DBM or MF-containing compositions due to the limited formation of brucite or other hydrates in such systems at low temperatures. In general, all evaluated formulations showed improved green mechanical strength and permeability levels when compared to the reference materials (AL-free formulations). As a consequence of the higher permeability level measured for the aluminum lactate-containing refractories, it was also possible to obtain $Al_2O_3$-MgO castables with greater spalling resistance even when they were subjected to a heating rate of 20°C/min. Therefore, the selected organic salt is an efficient additive to be used in the design of fast drying MgO-bonded monolithics.

## 5. Acknowledgments

The authors would like to thank Conselho Nacional de Desenvolvimento Cientifico e Tecnologico – CNPq (grant number: 303324/2019-8) and Fundação de Amparo a Pesquisa do Estado de São Paulo (FAPESP, grant number 2019/07996-0). The authors are also grateful to Almatis and RHI-Magnesita for supplying the raw materials used in this study.